\documentclass[12pt]{article}

\usepackage{scicite}

\usepackage{times}

\usepackage{graphicx}

\newenvironment{sciabstract}{%
\begin{quote} \bf}
{\end{quote}}

\newcounter{lastnote}
\newenvironment{scilastnote}{%
\setcounter{lastnote}{\value{enumiv}}%
\addtocounter{lastnote}{+1}%
\begin{list}%
{\arabic{lastnote}.}
{\setlength{\leftmargin}{.22in}}
{\setlength{\labelsep}{.5em}}}
{\end{list}}

\title{Observation of Rydberg blockade between two atoms }

\author
{E.  Urban, T. A. Johnson, T. Henage, L. Isenhower, D. D. Yavuz,\\ T. G. Walker, and M. Saffman$^\ast$\\
\\
\normalsize{Department of Physics, University of Wisconsin,}\\
\normalsize{1150 University Avenue, Madison, WI 53706 USA}\\
\\
\normalsize{$^\ast$To whom correspondence should be addressed; E-mail: msaffman@wisc.edu}
}

\date{}

\begin{document} 




\maketitle

\begin{sciabstract}
We demonstrate experimentally that a single Rb atom excited to the $79d_{5/2}$ level 
blocks the subsequent excitation of a second atom located 
more than $10~\mu\rm m$ away. 
The observed probability of double excitation of $\sim 30~\%$ is consistent  with a theoretical 
model based on  calculations of the long range dipole-dipole  interaction  between  atoms. 
\end{sciabstract}

Blockade interactions whereby  a single particle prevents the flow or excitation of other particles
 provide a generic mechanism for conditional control of  quantum states including entanglement of
 two or more particles. Blockade is  therefore of great interest for quantum information experiments
 and has  been observed for electrons (Coulomb\cite{ref.cb,ref.cb2} and spin blockade\cite{ref.sb}), photons\cite{ref.pb} 
 and also in trapping of cold atoms\cite{ref.grangier}.
Long range dipolar interactions between highly excited atoms  have been proposed as a mechanism for ``Rydberg  blockade"\cite{ref.jaksch2000,ref.lukin2001} 
which  enables a rich assortment of  quantum gate and entanglement protocols \cite{ref.gates1,ref.gates2,ref.gates3,ref.gates4}. 
Signatures of strong dipolar interactions between Rydberg atoms were first observed 
 several decades ago\cite{ref.haroche} and have recently been the 
subject of intense study using samples of cold atoms with which suppression of excitation due to Rydberg
 interactions has been observed in a many body 
regime\cite{ref.suppression1,ref.suppression2,ref.suppression3,ref.suppression4,ref.suppression5,ref.suppression6}.
In order to apply the Rydberg blockade effect to controlled evolution of quantum systems it is necessary
 to reach the regime of strong interactions between individual atoms.   

We report here on observation of 
Rydberg blockade   between 
single neutral atoms separated by more than $11~\mu\rm m$, which is an enabling step towards creation of
 entangled atomic states. Previous demonstrations of neutral atom entanglement have relied on short range
 collisions at length scales characterized by a 
low energy scattering length of about $10 ~\rm nm$\cite{ref.bloch,ref.phillips}.  Our results,  using laser
 cooled and optically trapped $^{87}$Rb, extend the distance for strong two-atom interactions by three orders
 of magnitude, and place us in a regime where the interaction distance is large
 compared to $1~\mu\rm m$ which is the  characteristic wavelength of light needed for internal state manipulation.
 The factor of ten we achieve between interaction length 
and wavelength of the control light is a significant step towards demonstration of a universal  quantum gate
  between neutral atoms in a scalable architecture.

\begin{figure}[!t]
\centering
\includegraphics[width=11.cm]{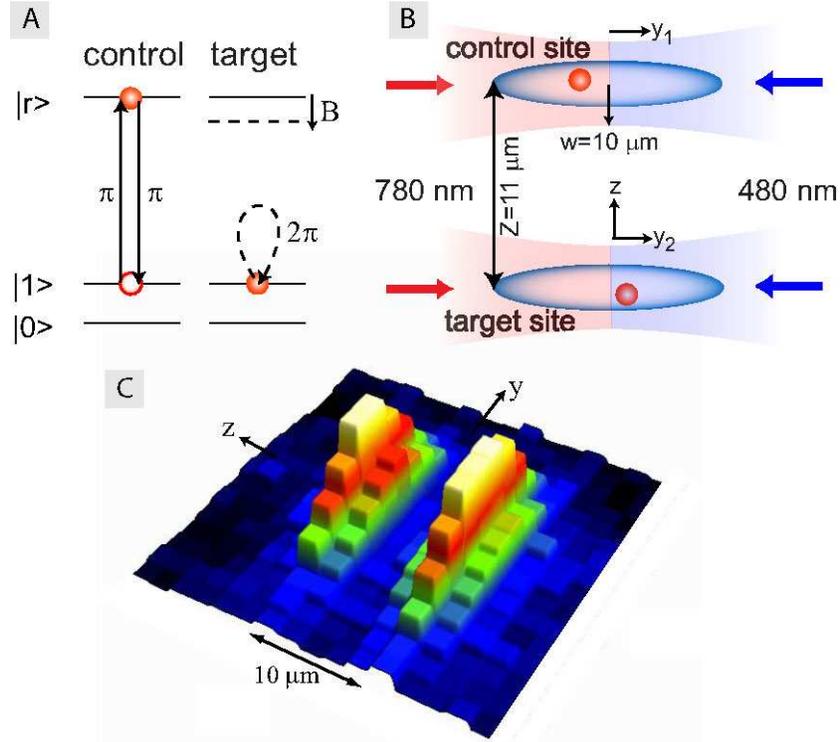}
\caption{A) Rydberg blockade between control and target atoms each with internal qubit states $|0\rangle, |1\rangle$ and Rydberg state $|r\rangle.$ A controlled phase operation on the target requires excitation of the control atom to $|r\rangle_{\rm c}$ with a $\pi$ pulse, a $2\pi$ pulse on the target atom, and a $\pi$ pulse on the control atom to return it to the ground state. When a control atom is initially present in $|1\rangle_{\rm c}$ it is excited to $|r\rangle_{\rm c}$ and the dipole-dipole  interaction  $|r\rangle_{\rm c}\leftrightarrow |r\rangle_{\rm t}$ shifts the Rydberg level by an amount ${\sf B}$ which detunes the excitation of the target atom so that it is blocked. The final $\pi$ pulse then returns the control atom to the ground state. B) Experimental geometry with two trapping regions separated by $Z=11~\mu\rm m$. 
States $|1\rangle, |r\rangle$ are coupled by two-photon transitions driven by counterpropagating 780 and 480 nm lasers polarized along $\hat z$ . Each laser is focused to a waist ($1/e^2$ intensity radius) of $w\sim 10~\mu\rm m$ and the light of each color is switched between control and target sites using acousto-optic modulators (AOMs).   C) Experimental fluorescence image of the atomic density  created by averaging 150 exposures each with one atom in the control and target sites.     }
\label{fig.idea}
\end{figure}

The mechanism of Rydberg blockade is shown in Fig. \ref{fig.idea}A.  Two atoms, one labeled ``control" and the other ``target" are placed in proximity with each other. The  ground state  $|1\rangle$ and Rydberg  
state $|r\rangle$  of each atom form a two-level system  that is coupled by laser beams with Rabi frequency $\Omega.$ Application of a 2$\pi$ pulse ($\Omega T=2\pi$ with $T$ the pulse duration) on the target atom results in excitation and de-excitation of the target atom giving  a phase shift of $\pi$ on the quantum state, i.e. $|1\rangle\rightarrow -|1\rangle.$   If the 
control atom is excited to the Rydberg state before application of the $2\pi$ pulse, the  excitation of the target is blocked and $|1\rangle\rightarrow|1\rangle$. Thus the excitation dynamics and phase of the target atom depend on the state of the control atom. Combining this Rydberg blockade mediated controlled-phase operation\cite{ref.jaksch2000} with $\pi/2$ single atom rotations between states $|0\rangle, |1\rangle$ of  the target  will implement the CNOT gate between two atoms. We have previously demonstrated the ability to perform  ground state rotations at individual trapping sites\cite{ref.yavuz2006}, as well as coherent excitation from ground to Rydberg states at a single site\cite{ref.johnson2008}. Here we describe two-site excitation experiments that demonstrate the Rydberg blockade  effect.

In order to perform the blockade operation of Fig. \ref{fig.idea}A the atoms must be close enough to have a strong interaction, yet far enough apart that they can be individually controlled (see Fig. \ref{fig.idea}B). To satisfy these conflicting requirements we first localize single atoms to regions that are formed by tightly focused beams from a far detuned laser (trapping wavelength $\lambda=1064~\rm nm$, waist $w=2.7~\mu\rm m$).  The far off resonance traps (FORTs)  have a potential depth of $U/k_B=5~\rm mK$ within which we trap atoms with temperature $T \simeq 150 ~\mu\rm K$
(the temperature is measured by releasing and recapturing single atoms). 
Based on the measured temperature we infer  position probability distributions for each atom that are quasi one-dimensional Gaussians with $\sigma_z=0.23~\mu\rm m$ and $\sigma_y=2.6~\mu\rm m$. 
The resulting single atom distributions observed experimentally by viewing along the $x$ axis are shown in  Fig. \ref{fig.idea}C.
  The lasers which control the coupling between internal states are focused to a small waist $w\sim 10~\mu\rm m$ so  the atoms can be separately manipulated by displacing the control lasers, even though the  trapping sites 
are  close together. In addition we excite a high lying Rydberg level with $n=79$. The strength of the long range interaction between two Rydberg atoms scales as $n^{11}$ with $n$ the principal quantum 
number\cite{ref.gallagherbook}. As will be  discussed in more detail below the  $79d$ Rydberg levels provide ${\sf B}/2\pi \simeq 2~\rm MHz$ of blockade shift at $Z=11~\mu\rm m$ which is sufficient for a strong two-atom blockade effect.

\begin{figure}[!t]
\centering
\includegraphics[width=11.cm]{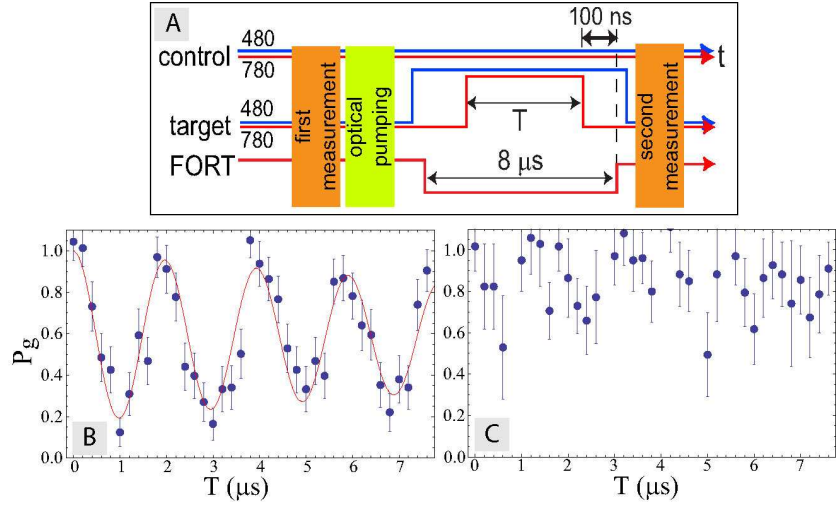}
\caption{\label{fig.crosstalk}Rabi flopping experiment to $79d_{5/2}$.  A)  
experimental sequence, B) measured ground state population during Rabi flopping on the target site and C)  crosstalk when the Rydberg excitation lasers are pointed at the empty control site. The Rabi pulse length $T$ is defined by a 780 nm laser which is switched with a fast AOM (20 ns rise time). The 480 nm light is switched more slowly, before and after each pulse.  
 Each data point is the average of $\sim30$ pre-selected single atom experiments, with the bars showing $\pm~ 1$ standard deviation.   The solid line shows a  curve fit to the data  with the function $(1-a)+a e^{-\frac{t}{\tau}}\cos(\Omega t)$ which gives a Rabi frequency of $\Omega=2\pi\times0.51~\rm MHz$, whereas our theoretical value with no adjustable parameters 
is\cite{rabicalc} $\Omega=2\pi\times 0.59~\rm MHz$.  We attribute the approximately 14\% lower experimental value to some spatial misalignment, and a small fraction of the Rydberg excitation light being present in servo sidebands from  laser locks.  The data show that the atom is returned to the ground state with  $95\%$ probability after one cycle and that Rydberg state excitation is achieved with $\sim 80 \%$ probability. The lack of perfect Rydberg excitation is due to several factors  which we estimate as 
Doppler broadening of the excitation($\sim 5-10\%)$, imperfect optical pumping ($\sim 5\%)$,  and imperfect detection efficiency ($\sim 5\%)$. }
\end{figure}

The experimental apparatus and procedures have been described   in our recent 
publication\cite{ref.johnson2008,ref.laser} where we observed coherent oscillations between the
$^{87}$Rb ground and Rydberg levels
 $|1\rangle\equiv |5s_{1/2},f=2,m_f=2\rangle$ and $|r\rangle\equiv |43d_{5/2},m_j=1/2\rangle$. Here we use similar techniques but  in order to get a sufficiently strong interaction
we now excite the $|r\rangle\equiv|79d_{5/2},m_j=1/2\rangle$ Rydberg level.  The essential steps of the experimental sequence 
 shown in Fig. \ref{fig.crosstalk}A are as follows. We start by loading single atoms into one or both sites from a cold background vapor in a magneto-optical trap. Both trapping sites are measured simultaneously by scattering red-detuned molasses light that is near resonant with the $5s_{1/2}-5p_{3/2}$ transition and imaging the fluorescence onto a sensitive CCD camera using imaging optics that give an effective pixel size of $2.1\times 2.1~\mu\rm m$. The counts in  regions of interest corresponding to the peaks in Fig. \ref{fig.idea}C
are tabulated and loading of a single atom is established
when the counts lie within upper and lower confidence bounds based on histograms of many 
measurements. The presence of atoms is checked by the first measurement, after which we optically pump the atoms into $|1\rangle.$ We then turn off the trapping potentials for $8~\mu\rm s$ while performing the Rydberg excitation, restore the trapping potential, and use a second measurement to see whether or not there is still an atom in each site. The trapping light photoionizes the Rydberg atoms faster than they can spontaneously decay to the ground state\cite{ref.swpra2005}, which enables detection of Rydberg excitation by monitoring atom loss.  Since the trap turn-off time is less than  the trap 
vibrational period (radial $ 12.3 ~\mu\rm s$ and axial $139~\mu\rm s$) the 
probability of atom loss due to turning off the trap is measured to be  a small $3\%.$

Rabi oscillations  between $|1\rangle$ and $|r\rangle$  are shown in Fig. \ref{fig.crosstalk}B when there is no atom present in the control site, and one atom in the target site.  The measured and calculated Rabi  frequencies agree to about 10\%.   Figure \ref{fig.crosstalk}C shows the crosstalk measured at the control site by loading one control atom, no target atoms, and keeping the lasers aligned to the target site.  The expected Rydberg 
population due to crosstalk given the Gaussian shape of our excitation lasers is
$P'\sim {\Omega'}^2/({\Omega'}^2 +\Delta_{ac}^2)$ where  
$\Omega'/\Omega=e^{- Z^2/w_{z,780}^2}e^{- Z^2/w_{z,480}^2}=0.019$ is the relative Rabi frequency at the non-addressed site and $\Delta_{\rm ac}=2\pi\times 2 ~\rm MHz$ is our theoretical estimate for the AC Stark shift of the $|1\rangle - |r\rangle$ transition. These numbers give $P'\sim 10^{-4}$   which is consistent with the observed lack of Rabi oscillations due to spatial crosstalk.

\begin{figure}[!t]
\centering
\includegraphics[width=11.cm]{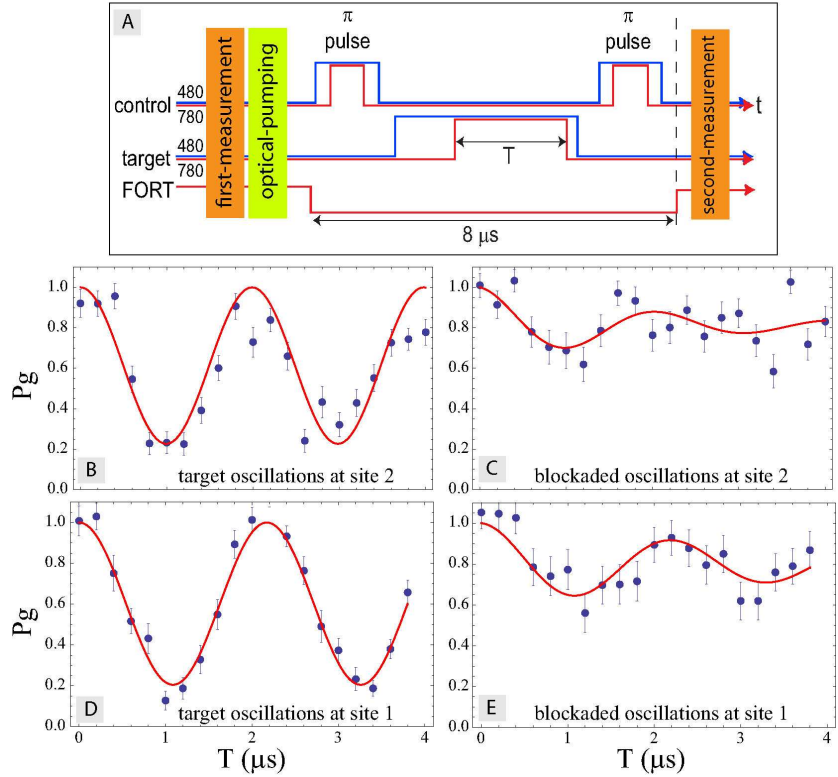}
\caption{\label{fig.blockade}Rydberg blockade experiment between control and target atoms. A) experimental sequence, B) Rabi oscillations on site 2 when no $\pi$ pulses are applied to site 1, C) blockaded oscillations on site 2 when the $\pi$  pulses are applied to site 1. Panels D) and E) show the same as B) and C) but with the roles of sites 1 and 2 reversed.      }
\end{figure}

Having verified the ability to perform Rabi oscillations between $|1\rangle$ and $|r\rangle$ we use the sequence shown in 
 Fig. \ref{fig.blockade}A to demonstrate Rydberg blockade. We start by loading one atom into each site and then perform a $\pi$ pulse $|1\rangle \rightarrow |r\rangle$ on the control site. A Rydberg pulse of length $T$ is then applied to the target, followed by a second $\pi$ pulse on the control atom to return it  to the ground state. 
We then measure the population in both sites and only keep the data for which we still have an atom in the control site. 
This procedure removes spurious data due to cases where the control atom  was lost during the initial  measurement. Rabi oscillations of the target with no control atom  (Fig. \ref{fig.blockade}B) can be compared with the data when there is a control atom present  (Fig. \ref{fig.blockade}C). We see that the presence of the control atom reduces the probability of populating the Rydberg state from about 0.8 to 0.3. 
We have verified that the experimental data are well reproduced by a stochastic rate equation model that accounts for random variations of atomic position, Doppler broadening, 
and the calculated blockade shift which is discussed below. 
The last two panels D and E show equivalent data but with the control and target roles switched between the two trapping sites.  
This verifies that the results are not due to some spatial asymmetry or bias in our experimental procedures.

\begin{figure}[!t]
\centering
\includegraphics[width=9.cm]{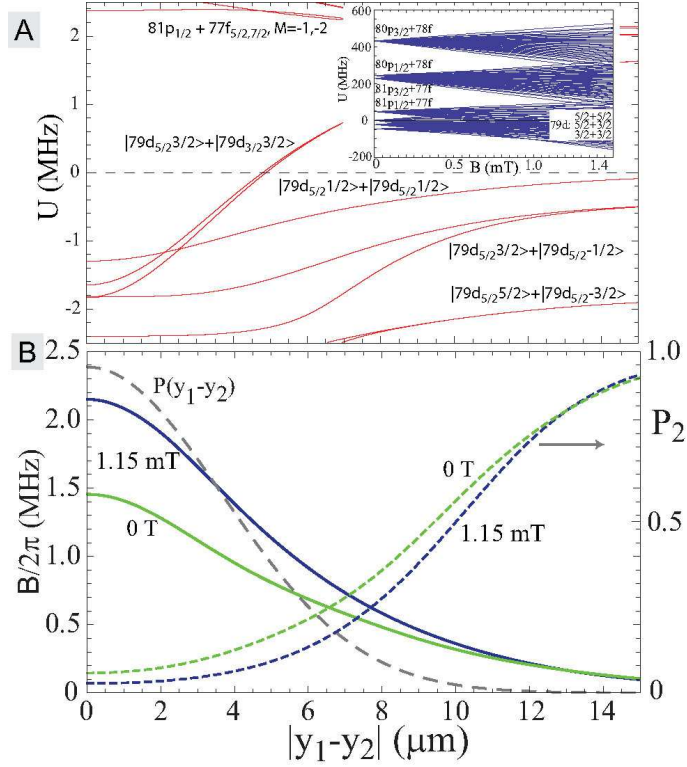}
\caption{\label{fig.theory} A) Molecular energies as a function of relative position $|y_1-y_2|$ with an external magnetic field of $1.15~\rm mT$. The inset shows the two-atom energies versus magnetic field for $R\rightarrow\infty$.  B) Blockade shift (solid lines) and $P_2$ (dashed lines) as a function of relative position $|y_1-y_2|$ for $0$ 
and $1.15~\rm mT$.  The long dashed line shows the 
relative probability of $y_1-y_2$ which is a Gaussian with variance $2\sigma_y^2$.
 }
\end{figure}

In order to explain these observations we must account quantitatively for the strength of the  Rydberg interactions. $^{87}$Rb atoms excited to the $79d_{5/2}$ state experience a F\"orster interaction\cite{ref.forster} that is dominated by the near degeneracy of the energy of two 79d atoms with the energy of a two atom state  $n_p p + n_f f$. The interaction is strongest for channels with\cite{ref.ws2008} $n_p=80, n_f=78$ and $n_p=81, n_f=77.$ The situation is complicated by the fact that the $B_0=1.15 ~\rm mT$ bias magnetic field 
that is used for optical pumping remains on during the Rydberg interaction giving  Zeeman shifts and coupling of different fine structure states.  This leads to mixing of the $79d_{5/2}$ and $79d_{3/2}$ fine structure manifolds which have a zero field separation of only about $23~\rm MHz.$ The laser excited Rydberg states are therefore linear superpositions of several $79d_{5/2}$ and $79d_{3/2}$ Zeeman states with $m=\pm 1/2, \pm 3/2.$ 
In addition the magnetic field differentially tunes the energies of the Zeeman sublevels which breaks the degeneracy leading to noninteracting 
 F\"orster zero states\cite{ref.ws2005}. The net result is that the interaction strength is substantially larger than it would be without a bias field. Two-atom energies for $R\rightarrow\infty$ as a function of
the  magnetic field are shown in the inset to Fig. \ref{fig.theory}A.
   We see that there is strong overlap of the $79d_{5/2} + 79d_{3/2}$  Zeeman fans  with $ 81p_{1/2}+77f_{5/2,7/2}$. We have calculated the interaction strength under these conditions by including all  436 two atom fine structure states taken from $(79d,79d)$, $(80p, 78f)$,  and   $(81p, 77f) $. 
The resulting $436\times 436$ Hamiltonian matrix including  energy offsets calculated from quantum defect 
data\cite{ref.gallagher,ref.gallagher2}, the Zeeman couplings, and the  dipole-dipole interaction, is then numerically diagonalized to find the molecular eigenvalues as a function of two-atom separation and relative orientation. Using all calculated molecular energies, a subset of which are shown in Fig. \ref{fig.theory}A,   we can characterize the interaction strength by a single number, the  blockade shift  ${\sf B}$, which determines the probability of double Rydberg excitation after a $\pi$ pulse on the target atom.

In the present experiment the probability distribution of $y_1-y_2$ extends to values where $P_2$ is not small compared to unity. In this situation the double excitation probability can be  approximated 
  by $P_2=\sum_\varphi \frac{\Omega^2\kappa_\varphi^2}{\Omega^2+2\Delta_\varphi^2}$,
 where $\kappa_\varphi^2=|\langle\varphi|rr\rangle|^2$ is the overlap between the laser excited two-atom  state $|rr\rangle$ and the molecular eigenstate $|\varphi\rangle$, and $\hbar \Delta_\varphi$ is the energy of the molecular states in Fig. 4A. Given $P_2$ we then calculate an effective blockade shift ${\sf B}$ 
using the definition $P_2=\frac{\Omega^2}{\Omega^2+2{\sf B}^2}$. 
The theoretical  blockade shifts and double excitation probabilities  for 0 and $1.15~\rm mT$ magnetic fields
found using the observed excitation Rabi frequency of $\Omega/2\pi=0.51~\rm MHz$  are shown in Fig. 4B. 
We see that the magnetic bias field serves to significantly strengthen the interaction relative to the zero field case. The  crossing of molecular eigencurves near $|y_1-y_2|=4.9~\mu\rm m$   with the $U=0$ line corresponding to the energy of the laser excited state  does not result in a noticeable perturbation of ${\sf B}$ 
since the  overlap at the  crossing point  is a very small $\kappa^2=|\langle 79d_{5/2}3/2,79d_{5/2}3/2|rr\rangle|^2\simeq 3\times 10^{-4}$.
Averaging over the probability distribution of $y_1-y_2$ we predict a double excitation probability of $\bar{P_2}= 0.069.$
This value of $\bar{P_2}$ in turn implies an effective blockade strength of 
$\bar{\sf B}=1.3~\rm MHz$.

 The data of Fig. \ref{fig.blockade} show double excitation probabilities in the blockaded case as high as $P_2 \sim 0.3 - 0.35$.  
This difference can  be explained by noting that the control atom only has a probability of about $0.8$ of being excited to the Rydberg level which increases the  predicted  $P_2$ to $P_2\sim 0.007 + (1-0.8)\times 0.8= 0.23$,  which is not far from  the experimentally observed value. We note that the experiments were performed at a characteristic  two-atom separation of $R\sim \sqrt{Z^2+2\sigma_y^2}=11.6~\mu\rm m.$ A modest improvement in focusing of the trapping and control lasers should allow working at $Z=7~\mu\rm m$ or better. Numerical calculations for this case predict $\bar{P_2}=0.007$ and further optimization should be possible by also improving the confinement along $y.$  
 Indeed a detailed analysis of a Rydberg blockade  CNOT gate taking into account a large number of experimental imperfections suggests that gate errors at the $10^{-3}$ level are feasible\cite{ref.swpra2005}.

In conclusion we have observed Rydberg blockade between two atoms localized in spatially separated trapping sites. The excitation of one atom to a Rydberg level blocks the subsequent excitation of a second atom. 
The blockade effectiveness is consistent with calculations of the Rydberg interaction strength accounting for all experimental parameters. Our observations complement previous work on excitation suppression in a many body regime\cite{ref.suppression1,ref.suppression2,ref.suppression3,ref.suppression4,ref.suppression5,ref.suppression6}  and extend the range of strong interactions between just two atoms to a distance that is ten times larger than the wavelength of the light needed for internal state manipulation. We believe this will be important for  ongoing efforts to build scalable neutral atom quantum logic devices. 
In future work we plan to combine the blockade studied here with single atom rotations in order to demonstrate a universal CNOT gate between two-atoms which may eventually form the basis for a many qubit quantum processor.

\bibliography{arxivR1}

\begin{thebibliography}{10}
\bibitem{rabicalc}
The theoretical two-photon Rabi frequency is $\Omega=C_\Delta
  \Omega_{780}\Omega_{480}/2\Delta.$ Starting from the $|f=2,m_f=2\rangle$
  ground state with $\hat z$ polarized light
  $\Omega_{780}=\frac{\sqrt2}{3}\frac{q{\mathcal E}_{780}}{\hbar}
  R_{5_{s_{1/2}}}^{5_{p_{3/2}}}$,
  $\Omega_{480}=\frac{\sqrt6}{5}\frac{q{\mathcal
  E}_{480}}{\hbar}R_{5_{p_{3/2}}}^{43_{d_{5/2}}}$, and
  $C_\Delta=(\Delta-\Delta_{23}/2)/(\Delta-\Delta_{23})$ with
  $\Delta_{23}/2\pi=-0.267~\rm GHz$ the hyperfine splitting of the $5p_{3/2}$
  $f=2,3$ levels, and $\Delta$ the laser detuning from $|5p_{3/2},f=3\rangle$.
  Using quantum defect wavefunctions we find
  $R_{5_{s_{1/2}}}^{5_{p_{3/2}}}=5.14a_0, R_{5_{p_{3/2}}}^{43_{d_{5/2}}}=
  0.012a_0.$ Experimental parameters of $\Delta/2\pi=-2.4~\rm GHz$, beam powers
  of $P_{780}=1.95~\mu\rm W,$ $P_{480}=26.2~\rm mW,$ and waists $w_{x,
  780}=11.9,$ $ w_{z, 780}=7.0,$ $ w_{x, 480}=11.3, w_{z, 480}=9.0~\mu\rm m$
  give $\Omega/2\pi=0.61~\rm MHz.$ This is then reduced by $3.7\%$ to account
  for the Zeeman mixing of the $79d_{5/2}$ state with $79d_{3/2}$ which has a
  negligible radial matrix element with $5p_{3/2}.$ The final result is the
  quoted value of $\Omega/2\pi = 0.59~\rm MHz.$.
\bibitem{ref.laser}
In the present work we use a slightly different optical trapping wavelength,
  1064 nm light generated by a fiber laser instead of the previously used 1030
  nm. The change was made due to an equipment failure and is not crucial to the
  experimental results.
\end{thebibliography}

\bibliographystyle{Science}

\begin{scilastnote}
\item 
This work was supported by the NSF and ARO-IARPA. 
\end{scilastnote}

\end{document}